\documentclass[twocolumn]{aastex7}
\usepackage{multirow}

\hypersetup{colorlinks,%
 citecolor=blue,%
 linkcolor=blue}
\newcommand{\kms}{km\,s$^{-1}${}}
\newcommand{\msun}{M$_{\sun}${}}
\shorttitle{Groups of Dwarf Galaxies}
\shortauthors{Paudel et al.}

\newcommand{\trev}[1]{{\textbf{#1}}}

\newcommand{\tre}[1]{{\color{red}{#1}}}
\newcommand{\tbl}[1]{{\color{blue}{#1}}}
\newcommand{\tgr}[1]{{\color{green}{#1}}}

\begin{document}

\title{Groups of Dwarf Galaxies in the Local Universe }

\author[orcid=0000-0003-2922-6866,gname='Sanjaya',sname='Paudel']{Sanjaya Paudel}
\affiliation{Department of Astronomy \& Center for Galaxy Evolution Research, Yonsei University, Seoul 03722, Republic Of Korea }
\email{sanjpaudel@gmail.com}  

\author[orcid=0000-0002-5513-5303,gname='Cristiano G.', sname='Sabiu']{Cristiano G. Sabiu} 
\affiliation{Natural Science Research Institute (NSRI), University of Seoul, Seoul 02504, Republic of Korea}
\email{csabiu@gmail.com}

\author[orcid=0000-0002-1842-4325,sname='Yoon']{Suk-Jin Yoon}
\affiliation{ Department of Astronomy \& Center for Galaxy Evolution Research, Yonsei University, Seoul 03722, Republic Of Korea}
\email[show]{sjyoon0691@yonsei.ac.kr}

\author[orcid=0000-0002-6841-8329,sname='Chhatkuli']{Daya Nidhi Chhatkuli}
\affiliation{Department of Physics, Tri-Chandra Multiple Campus, Tribhuvan University, Kathmandu, Nepal}
\email{chhatkulidn@gmail.com}

\author[orcid=0000-0002-6841-8329,sname='Yoo']{Jaewon Yoo}
\affiliation{Korea Astronomy and Space Science Institute, Daejeon 34055, Republic of Korea}
\email{jwyoo@kasi.re.kr}

\author[orcid=0000-0003-2569-8129,sname='Pokhrel']{Nau Raj Pokhrel}
\affiliation{Department of Physics and Astronomy, The University of Tennessee, Knoxville, TN 37996, USA}
\email{npokhrel@utk.edu}


\begin{abstract}

We present a systematic search for dwarf-galaxy groups in the local Universe ($z<0.02$), identifying 28 systems containing at least four spectroscopically confirmed members selected from the SDSS and DESI surveys. Group membership is assigned based on projected separation -within 300\,kpc from the most massive galaxy (designated the ``central'') -and a relative line-of-sight velocity difference of less than 200\,\kms. 
The sample has a median redshift of $z=0.0131$ ($\sim55$\,Mpc) and a median central stellar mass of $1.7\times10^{9}\,M_\odot$, firmly placing these systems in the low-mass regime. 
In total, 28 groups contain 129 dwarf galaxy members, with a median stellar mass of $1.63\times10^{8}\,M_\odot$. The median $r$-band apparent magnitude of the member dwarfs is 17.38\,mag, slightly fainter than the spectroscopic observation limit of the SDSS main spectroscopic survey. Remarkably, several groups are centered on galaxies with stellar masses as low as $\sim10^{8}\,M_\odot$, comparable to the Fornax dwarf spheroidal, demonstrating that even very faint galaxies can host bound satellite systems. Most groups exhibit low velocity dispersions ($\sigma_{v}<50$\,\kms), consistent with being gravitationally bound. 
The inferred dynamical masses span $\sim10^{10}-10^{12}\,M_\odot$, while the corresponding three-dimensional velocity dispersions ($\sigma_{3D}$) fall between 50 and 100\,\kms, characteristic of dynamically cold, low-mass halos. Our results provide empirical constraints on small-scale structure formation and show strong consistency with predictions from cosmological volume simulations, supporting the picture that dwarf galaxies can serve as central hosts for their own satellite systems embedded within extended dark-matter halos.
\end{abstract}

\section{Introduction}

The Lambda Cold Dark Matter ($\Lambda$CDM) model is the prevailing cosmological framework for understanding the formation and evolution of the Universe's large-scale structure. It successfully accounts for a wide range of observational phenomena, including the cosmic microwave background, galaxy clustering, and the distribution of dark matter halos on scales exceeding tens of megaparsecs \citep{White78, Frenk12}. However, despite its successes on large scales, the $\Lambda$CDM paradigm encounters significant challenges when applied to the properties and dynamics of dwarf galaxies --systems with stellar masses typically below $10^{9}\,M_\odot$ that serve as critical tests of the model in the low-mass regime \citep{Bullock17}.

Within the $\Lambda$CDM framework, structure formation proceeds hierarchically through the gravitational collapse of dark matter halos. A key prediction is that even low-mass halos, with virial masses of $10^{9}$--$10^{11}\,M_\odot$, should host substantial substructure due to the shape of the primordial power spectrum \citep{Diemand08}. These subhalos are expected to harbor luminous dwarf galaxies, implying that dwarfs should rarely exist in complete isolation but instead may host their own populations of faint satellites or assemble into gravitationally bound associations \citep{Kravtsov10, Sales13, Wheeler15, Yaryura20}.

\begin{figure*}
\includegraphics[width=17.5cm]{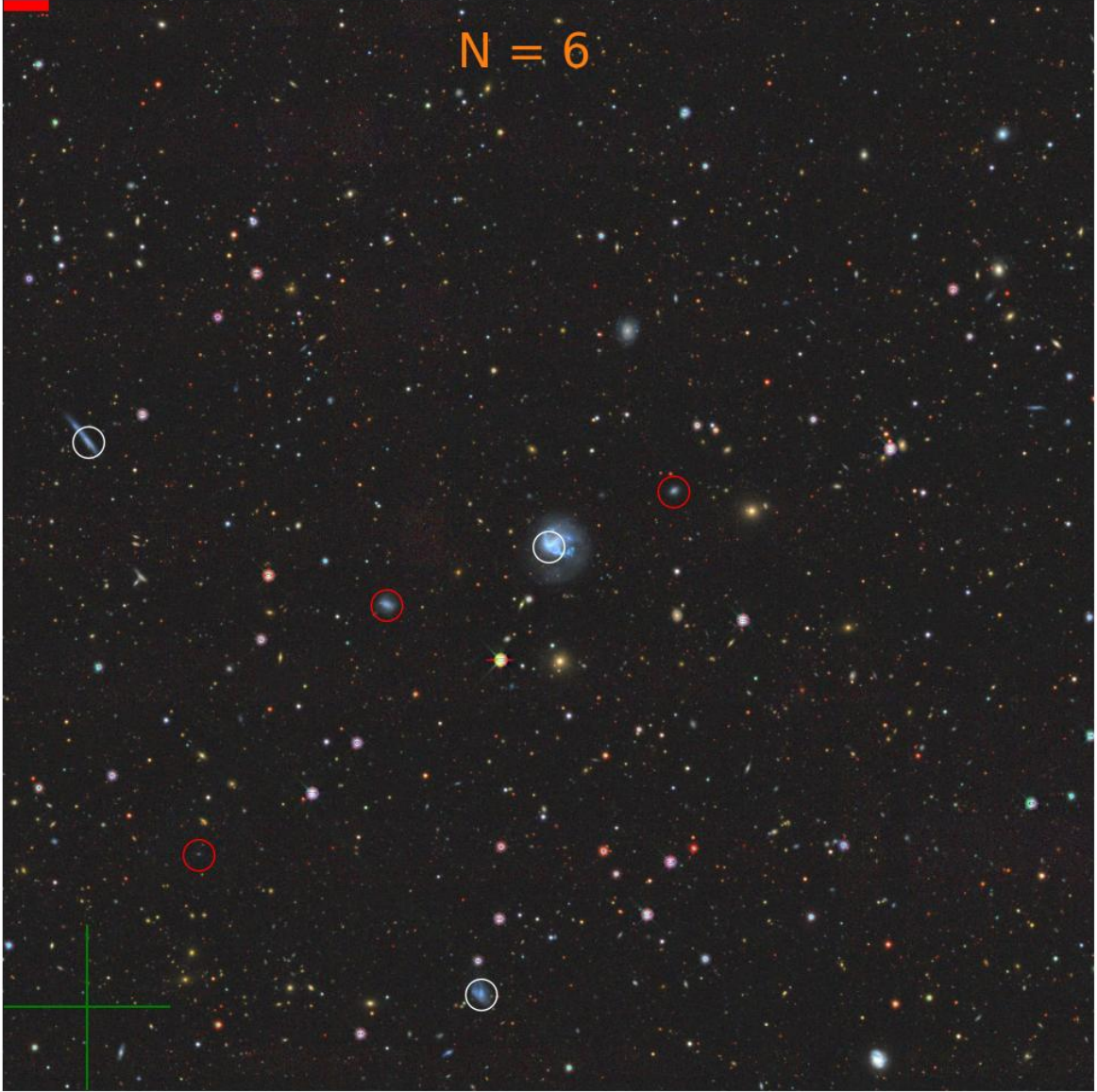}
\caption{Example of a dwarf galaxy group (ID = 0948$-$034) consisting of six member galaxies. The positions of the member galaxies are marked with circles; galaxies with H\,{\sc i} measurements are shown in white. The cross in the lower-left corner indicates the $\sim3\arcmin$ diameter beam size typical of single-dish H\,{\sc i} observations such as ALFALFA and FASHI. A thick red horizontal line, shown in the top-left corner, represents a scale of 10 kpc.
The central galaxy of the group is NGC 4639, which has a stellar mass of 1.2$\times$10$^{9}$\,\msun. The background color image is obtained from the Legacy Survey SkyView server, covering a field of view of 20${\arcmin}$$\times$20${\arcmin}$. All remaining group pictures are shown in Appendix.}
\label{lgrp}
\end{figure*}


High-resolution cosmological simulations based on $\Lambda$CDM therefore predict a significant frequency of merger events involving dwarf galaxies alone \citep{Deason14, Wetzel15, Martin21}. These merger rates are expected to peak at high redshift (e.g., $z\sim2$--6), when the Universe was denser and interactions more frequent, and to continue at a reduced level to the present day. Such predictions follow from the steep subhalo mass function ($dN/dM \propto M^{-1.9}$) \citep{Springel08}, which ensures a large reservoir of low-mass systems capable of interacting and merging over cosmic time.

\begin{table*}
\caption{Physical properties of our dwarf galaxy groups }
\centering\
\setlength{\tabcolsep}{3pt}
\scriptsize
\begin{tabular}{lcccccccccccc|cc}
\hline
No. & ID & R.A. & Dec. & z& D & N & R & $\sigma_{v}$ & M$_{dyn}$ & M$_{bar}$ & M$_{H\,{\sc i} }$ & $\sigma_{3d}$  & M$_{*}$ & $M_{B}$  \\
 & $hhmmddmm$ & deg & deg &  &Mpc &   & kpc & \kms & log(\msun) & log(\msun) & log(\msun) & \kms & log(\msun) & mag\\
 1 & 2 & 3 & 4 & 5& 6& 7&8&9&10&11&12&13&14 & 15\\
\hline
01 &  0017+0643  & 004.3298 &   06.7253 &  0.0186 &  79.3 & 5 &  184.08 & 55.1 & 11.3 & 10.31 & 09.98(2) &  98.41 &  9.96 & $-$19.34\\
02 &  0125+0759  & 021.3958 &   07.9908 &  0.0097 &  41.4 & 4 &  048.05 & 52.0 & 10.9 & 09.71 & 09.69(2) & 104.09 &  8.20 & $-$16.73\\
03 &  0426-0440  & 66.55425 & $-$4.6813 &  0.0114 &  48.7 & 5 &  164.64 & 25.7 & 10.9 & 09.60 & 09.51(3) &  41.06 &  8.66 & $-$15.98\\
04 &  0728+3532  & 112.2018 &   35.5480 &  0.0131 &  55.9 & 4 &  072.40 & 38.8 & 10.9 & 10.04 & 09.77(1) &  76.32 &  9.62 & $-$18.85\\
05 &  0729+3341  & 112.4331 &   33.6899 &  0.0160 &  68.2 & 6 &  138.27 & 53.9 & 11.3 & 10.43 & 10.25(2) & 102.06 &  9.88 & $-$19.13\\
06 &  0803+0841  & 120.9851 &   08.6995 &  0.0165 &  70.4 & 4 &  103.69 & 96.0 & 11.9 & 10.50 & 10.15(2) & 152.06 &  10.0 & $-$19.51\\
07 &  0832+6622  & 128.1960 &   66.3702 &  0.0176 &  75.0 & 4 &  056.68 & 23.3 & 10.4 & 09.53 &$---$  &  46.18 &  8.71 & $-$17.06\\
08 &  0833+2932  & 128.3454 &   29.5383 &  0.0069 &  29.5 & 4 &  117.46 & 37.7 & 11.2 & 10.11 & 09.70(2) &  72.40 &  9.84 & $-$19.17\\
09 &  0847+1002  & 131.8238 &   10.0423 &  0.0109 &  46.5 & 4 &  035.87 & 30.1 & 10.0 & 09.01 & 08.88(1) &  57.99 &  8.03 & $-$16.11\\
10 &  0851$-$0221 & 132.9056& $-$2.3619 &  0.0111 &  47.4 & 7 &  171.45 & 30.4 & 11.1 & 10.21 & 09.92(1) &  55.90 &  9.66 & $-$18.76\\
11 &  0902+1306  & 135.6711 &   13.1092 &  0.0168 &  71.6 & 4 &  166.26 & 59.2 & 11.1 & 10.78 & 10.05(3) & 110.95 &  9.68 & $-$19.06\\
12 &  0942+0929  & 145.7211 &   09.4917 &  0.0107 &  45.7 & 6 &  191.55 & 32.6 & 10.9 & 10.22 & 10.14(2) &  60.57 &  9.16 & $-$18.54\\
13 &  0946+0542  & 146.5794 &   05.7090 &  0.0097 &  41.4 & 4 &  135.02 & 114. & 11.6 & 09.81 & 09.68(1) & 207.90 &  8.56 & $-$17.19\\
14 &  0948$-$0344 & 147.0802 &$-$3.7344 &  0.0129 &  55.0 & 6 &  139.30 & 61.4 & 11.8 & 10.07 & 09.84(3) &  91.13 &  9.55 & $-$18.85\\
15 &  0951+3035  & 147.7593 &   30.5843 &  0.0147 &  62.7 & 5 &  092.01 & 34.7 & 11.3 & 10.12 & 09.83(2) &  67.20 &  9.60 & $-$18.42\\
16 &  1017+0419  & 154.2988 &   04.3310 &  0.0045 &  19.2 & 4 &  073.54 & 11.2 & 10.0 & 08.76 & 08.57(1) &  13.81 &  8.11 & $-$15.47\\
17 &  1024+5723  & 156.1092 &   57.3888 &  0.0083 &  35.4 & 4 &  144.29 & 23.1 & 10.9 & 09.57 & 09.41(2) &   5.02 &  8.86 & $-$17.47\\
18 &  1127+5555  & 171.9224 &   55.9202 &  0.0185 &  78.9 & 4 &  084.80 & 39.0 & 11.4 & 10.01 & 09.49(1) &  66.83 &  9.50 & $-$18.38\\
19 &  1244+6214  & 191.0503 &   62.2474 &  0.0087 &  37.1 & 5 &  129.10 & 31.8 & 10.8 & 09.64 & 09.37(3) &  49.23 &  8.91 & $-$16.84\\
20 &  1413+1401  & 213.3207 &   14.0270 &  0.0145 &  61.9 & 4 &  059.73 & 42.6 & 10.8 & 09.61 & 09.55(1) &  85.16 &  8.26 & $-$16.46\\
21 &  1429+4426  & 217.4633 &   44.4475 &  0.0092 &  39.3 & 6 &  130.61 & 55.3 & 11.7 & 09.93 & 09.56(1) &  97.67 &  9.08 & $-$17.67\\
22 &  1431+2714  & 217.7869 &   27.2367 &  0.0150 &  64.0 & 4 &  069.74 & 79.4 & 11.3 & 09.97 & 09.74(1) & 115.02 &  9.56 & $-$19.23\\
23 &  1553+2059  & 238.3781 &   20.9837 &  0.0185 &  78.9 & 4 &  110.38 & 34.1 & 11.4 & 09.91 & 09.77(1) &  68.24 &  9.22 & $-$18.24\\
24 &  1605+4120  & 241.4415 &   41.3449 &  0.0066 &  28.2 & 5 &  091.80 & 36.8 & 11.0 & 09.77 & 09.60(2) &  70.82 &  9.22 & $-$18.31\\
25 &  1738+5153  & 264.5359 &   51.8974 &  0.0192 &  81.8 & 5 &  167.94 & 56.9 & 11.9 & 09.03 & $---$ &  79.38 &  8.74 & $-$16.74\\
26 &  2118+0215  & 319.7105 &   02.2620 &  0.0181 &  77.2 & 4 &  022.44 & 72.0 & 11.2 & 09.99 & 09.64(1) & 139.37 &  9.72 & $-$18.92\\
27 &  2212+0030  & 333.1576 &   00.5113 &  0.0139 &  59.3 & 4 &  021.38 & 55.3 & 10.7 & 09.71 & 09.26(1) & 107.31 &  9.50 & $-$16.70\\
28 &  2339+1052  & 354.9117 &   10.8689 &  0.0167 &  71.2 & 4 &  097.18 & 12.9 & 09.9 & 09.69 & 09.39(1) &  22.65 &  9.26 & $-$18.19\\
\hline
\end{tabular}
\tablecomments{The second column lists the group ID in hhmmddmm format. Columns 3, 4, and 5 provide the central galaxy’s coordinates and redshift. Column 6 shows the group’s distance, calculated from the Hubble flow assuming a cosmology with $\Omega_{m} = 0.27$ and $H_{0} = 71$ \kms. Columns 7-13 list the group richness (number of member galaxies), physical size (projected distance from the central to the most distant satellite), velocity dispersion, dynamical mass, total baryonic mass, total HI mass (number of members that have HI measurement ) and three-dimensional velocity dispersion ($\sigma_{3d}$), respectively. In the last two columns, we list properties of the central galaxy, such as its stellar mass and $B$-band absolute magnitude.
}
\label{mtb}
\end{table*}

Observationally, the TiNy Titans (TNT) Survey \citep{Stierwalt15, Stierwalt17} has shown that isolated dwarf galaxies exhibit a low degree of clustering, suggesting that most field dwarfs reside in relative isolation. A detailed study of dwarf–dwarf interactions using TNT data and cosmological simulations finds that fewer than 5\% of dwarf galaxies possess close companions---defined as neighbors within 50--100 kpc and with relative velocity differences below $100\,\mathrm{km\,s^{-1}}$ \citep{Besla18}. This indicates that tightly bound dwarf pairs are uncommon in low-density environments.

In contrast, studies of the satellite system around the Large Magellanic Cloud (LMC) reveal a markedly different picture. Several works have reported a high concentration of faint dwarf satellites within $\sim$50 kpc of the LMC, consistent with the presence of a gravitationally bound satellite group \citep{Onghia08, Koposov15, Jethwa16, Dooley17, Sales17}. These findings suggest that the LMC may have brought a cohort of companions during its infall into the Milky Way halo. Supporting this scenario, results from the FIRE (Feedback in Realistic Environments) simulations show that LMC-mass galaxies can host 5--15 bound dwarf companions with steep radial number-density profiles \citep{Hopkins14, Wetzel16}. These simulations reinforce the idea that dwarf galaxy groups or associations can form and persist as coherent structures before being accreted into more massive halos such as that of the Milky Way.

\begin{table*}
\caption{Properties of member galaxies}
\begin{tabular}{ccccccccc}
\hline
Group ID  & R.A. & Dec. & z & M$_{B}$ & $g$ & $r$ & M$_{*}$ &  M$_{H\,{\sc i} }$  \\
 hhmmddmm& deg & deg & & mag & mag & mag & log(\msun) & log(\msun) \\
  1 & 2 & 3 &4 & 5 &6 & 7 & 8 & 9 \\
\hline
0017+0643 &   4.3299  &  6.7253 &  0.01869 &     $-$19.35 & 14.80 & 14.36 &  9.96 &  9.79 (a)  \\  
0017+0643 &   4.2914  &  6.7196 &  0.01902 &     $-$17.67 & 16.51 & 16.16 &  9.13 &  9.55 (a)  \\   
0426$-$0440 &  66.5543  & $-$4.6814 &  0.01148 & $-$15.99 & 17.10 & 16.63 &  8.66 &  9.2 (b)  \\  
0426$-$0440 &  66.5656  & $-$4.6108 &  0.01130 & $-$14.46 & 18.65 & 18.27 &  7.90 &  ---   \\
0426$-$0440 &  66.6214  & $-$4.6702 &  0.01125 &  $-$15.17 & 17.95 & 17.60 &  8.13 &  8.74 (b)  \\  
1024+5723 & 156.1092  & 57.3888 &  0.00833 &     $-$17.47 & 14.99 & 14.75 &  8.87 &  9.33 (c)  \\  
1024+5723 & 156.2538  & 57.3583 &  0.00816 &     $-$14.11 & 18.44 & 18.48 &  7.04 &  ---   \\
1429+4426 & 217.4633  & 44.4476 &  0.00922 &     $-$17.68 & 14.98 & 14.66 &  9.08 &  9.56 (d)  \\  
1429+4426 & 217.4448  & 44.4363 &  0.00923 &     $-$13.92 & 18.81 & 18.71 &  7.20 &  ----   \\
\hline
\end{tabular}
\tablecomments{  Column (1): Group ID according to Table \ref{mtb}, Column (2): R.A., Column (3): Dec., Column (4): redshift, Column (5): $B$-bang absolute magnitudes,
Column (6,7): $g$ and $r$-band magnitude, Column (8): Stellar mass, Column (9): HI mass, sources are in bracket a: \citep{Haynes18} , b: \citep{Zhang24}, c: \citep{Paturel03}, d: \citep{Schneider92}
(This table shows a representative subset of the data; the full table is available in machine-readable form.)
}
\label{mrt}
\end{table*}

Despite detailed studies of several intriguing individual systems \citep{Bellazzini13, Koposov15, Stierwalt15, Paudel24, Freitas24, Correnti25}, it remains unclear whether these examples are representative of the overall dwarf population. The statistical properties of dwarf galaxy clustering---particularly beyond the immediate vicinity of massive hosts---are still poorly constrained due to the scarcity of systematic observational samples. Yet this represents a key gap, as dwarf galaxies constitute the majority of all galaxies in the Universe, and their interactions, mergers, and groupings likely play a significant role in shaping their evolution.

In this work, we present an observational census of dwarf galaxy groups and characterize their physical and dynamical properties. These measurements provide empirical constraints on the abundance and dynamical state of low-mass galaxy systems and enable direct comparisons with cosmological simulations, improving our understanding of galaxy evolution in the low-mass regime where baryonic processes strongly influence dynamical evolution.

\section{Sample Selection}

\subsection{Selection of Dwarf Galaxy Groups}
As part of our broader effort to understand the interaction-driven evolution and clustering behavior of low-mass galaxies, our primary goal is to construct a comprehensive catalog of dwarf galaxy groups containing at least four member galaxies. This work extends our earlier studies of interacting dwarf–dwarf pairs, originally compiled to identify potential dwarf galaxy mergers. Building on that foundation, we subsequently identified several interacting dwarf pairs that host additional nearby members within the gravitational influence of a massive galaxy, and for a subset of these systems, we have already conducted detailed case studies \citep{Paudel24}.

To expand the sample and examine the frequency and properties of more complex dwarf associations, we performed a systematic search for additional dwarf companions around known dwarf pairs. This search utilized spectroscopic redshift measurements from the Sloan Digital Sky Survey \citep[SDSS\footnote{We used DR8 spectroscopic catalog which are obtained from applying a querry in the SDSS sky server  i.e, https://skyserver.sdss.org/casjobs/};][]{Aihara11} and the Dark Energy Spectroscopic Instrument Data Release~1 \citep[DESI DR1\footnote{\url{https://data.desi.lbl.gov/}};][]{DESI25}. We identified physically associated systems using both spatial and kinematic criteria: dwarf galaxies were selected if they lie within a projected radius of 300~kpc of the most massive group member (hereafter the ``central'') and have relative line-of-sight velocity differences smaller than 200~\kms. These criteria are intended to isolate galaxies that are likely gravitationally bound and dynamically interacting within the group potential. The basic properties of the groups are listed in Table~\ref{mtb}, and an illustrative example of a six-member group is shown in Figure~\ref{lgrp}.

To ensure statistical robustness and minimize contamination from chance projections, we restrict our final sample to systems with at least four spectroscopically confirmed members. Dwarf galaxies are selected from the combined spectroscopic catalogs of SDSS and DESI DR1. Although DESI reaches fainter magnitudes than SDSS, its current footprint and targeting are not yet as homogeneous as those of the SDSS Main Galaxy Sample. For this reason, and to quantify observational incompleteness in a uniform manner, we adopt the well-defined SDSS spectroscopic flux limit as our reference selection function.

Following \citet{Besla18}, we compute, at each galaxy redshift, the stellar mass corresponding to the SDSS Main Galaxy Sample limit ($m_{r} < 17.77$ mag). This procedure yields a mass-dependent completeness curve, from which we determine that the sample is $\geq$90\% complete above $\log(M_*/M_\odot)=8.63$. We adopt this stellar-mass threshold in all subsequent statistical analyses and simulation comparisons.

The presence of group members with inferred stellar masses approaching the SDSS sensitivity limit indicates that the census becomes increasingly incomplete at lower masses. Consequently, the derived dwarf-group fraction should be interpreted as a conservative lower bound if the definition of ``dwarf'' were extended to smaller stellar masses.


\subsection{Data Analysis}

This study leverages extensive multi-wavelength public datasets, allowing robust measurements of the physical and dynamical properties of the groups and their member galaxies. Spectroscopic redshifts from DESI and SDSS were used to confirm group membership via radial velocities. Optical $g$, $r$, and $z$-band magnitudes were obtained from the Legacy Survey imaging database\footnote{We used ls\_dr10.tractor catalog.}, and additional catalog information was retrieved from the Strasbourg Astronomical Data Center (CDS)\footnote{\url{https://cds.unistra.fr}}, including 21-cm H\,{\sc i}  fluxes used to estimate neutral hydrogen masses.

To derive stellar masses and $B$-band magnitudes, we primarily use the imaging catalog provided by the Legacy Survey. In most cases, the member galaxies are well separated, allowing us to adopt the catalog photometry directly. However, for a few systems where member galaxies are closely separated or exhibit diffuse morphologies, we perform our own aperture photometry to measure $g$- and $r$-band magnitudes. For this purpose, we use archival images from the Legacy Survey and follow the methodology described in our main catalog paper \citep{Paudel18}. 

We estimate $B$-band magnitudes from the measured $g$ and $r$ photometry using the transformation
\[
B = g + 0.227(g-r) - 0.337,
\]
\footnote{\url{https://www.sdss3.org/dr8/algorithms/sdssUBVRITransform.php}}
and derive stellar masses from the $r$-band luminosities using the mass-to-light ratio calibration from \citet{Zhang17}. The resulting photometric measurements are summarized in Table~\ref{mrt}.

Fortunately, most of the sky regions containing our dwarf galaxy groups are covered by the Five-hundred-meter Aperture Spherical Radio Telescope (FAST\footnote{\url{https://fast.bao.ac.cn}}), the FAST All Sky H\,{\sc i}  Survey \citep[FASHI;][]{Zhang24}, or the Arecibo Legacy Fast ALFA Survey \citep[ALFALFA;][]{Haynes18}. For a few galaxies, we also found an H\,{\sc i} archival catalog provided by CDS, which particularly comes from \cite{Schneider92,Paturel03}. We extracted H\,{\sc i} 21-cm flux measurements from these surveys to estimate the neutral hydrogen content of the groups. Because these archival  H\,{\sc i}  are from single-dish surveys with typical beam sizes of $\sim3\arcmin$, the detected flux may include contributions from multiple galaxies within a group. Consequently, in several cases, the measured flux represents the total H\,{\sc i} mass of the system rather than that of individual galaxies. In Table \ref{mtb} (column 12) we list the total H\,{\sc i} mass for each group, with the number of member dwarf galaxies that have individual H\,{\sc i} measurements indicated in parentheses. The H\,{\sc i} measurements for individual galaxies are also reported in Table \ref{mrt}.

\section{Result}
\begin{figure}

\includegraphics[width=8.5cm]{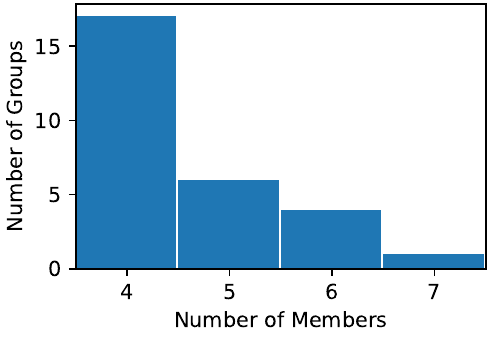}
\caption{Richness distribution of our dwarf galaxy groups, measured in the number of members.}
\label{ndis}
\end{figure}

\begin{figure}
\includegraphics[width=8.5cm]{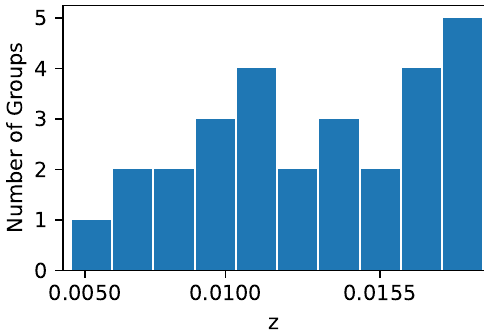}
\includegraphics[width=8.5cm]{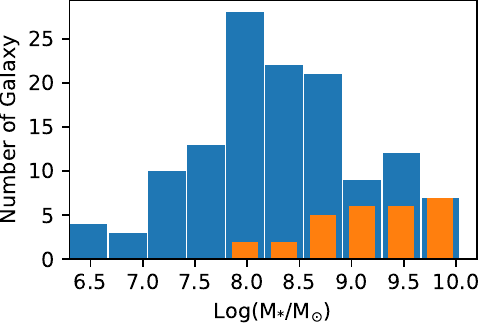}
\caption{Top: Redshift distribution of our group of dwarf galaxies. Bottom: Distribution of the logarithm of the stellar mass of member galaxies (in blue) and the central galaxy of the group  (in orange).}
\label{zdis}
\end{figure}

\begin{figure}
\includegraphics[width=8.5cm]{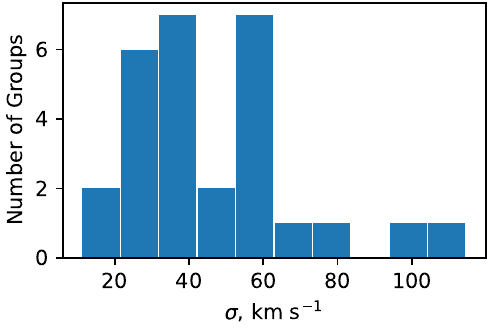}
\includegraphics[width=8.5cm]{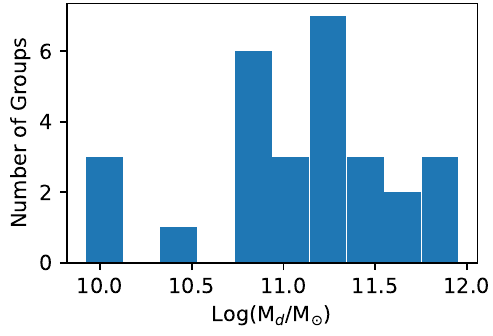}
\caption{Top: Distribution of standard deviation ($\sigma$) of line-of-sight velocities of our dwarf galaxy groups. \\
Bottom: Distribution of the logarithm of the dynamical mass of the dwarf galaxy groups. The dynamical mass is calculated using the projected mass formula provided by \cite{Heisler85}.}
\label{sighist}
\end{figure}

Figure~\ref{ndis} shows the richness distribution of our dwarf galaxy groups, measured by the number of identified members. As expected, the largest fraction of systems contains four galaxies, while the richest group in our sample includes seven members.

In Figure~\ref{zdis}, we present the redshift and stellar mass distributions of the central galaxies. The median redshift of the sample is $z=0.0131$ (corresponding to a comoving distance of $\sim 55$~Mpc), placing the groups firmly within the local Universe. The median stellar mass of the central galaxies is $1.7\times10^9$\,\msun, consistent with the low-mass regime characteristic of dwarf systems. Notably, our sample includes groups whose central galaxies have stellar masses as low as $\sim10^8$\,\msun, comparable to that of the Fornax dwarf spheroidal galaxy, one of the well-known satellite galaxies of the Milky Way. This highlights the depth of our search and demonstrates that even very low-mass galaxies can serve as central objects within small-scale gravitationally bound systems.

Figure~\ref{sighist} displays the distribution of line-of-sight velocity dispersions ($\sigma_{v}$; top panel) and the corresponding dynamical mass estimates (bottom panel). The majority of the systems exhibit relatively low velocity dispersions, with $\sigma_{v} < 50$\,\kms, strongly suggesting that these systems are gravitationally bound. This low velocity dispersion is consistent with expectations for dynamically cold systems such as dwarf groups, where internal kinematics is dominated by mutual gravitational interactions rather than random motions.

To estimate the total dynamical mass of each group, we employed the projected mass estimator of \citet{Heisler85}:
$M_{d} =\frac{32}{\pi}\frac{1}{G(N-3/2)} \sum\limits_{i}^{N} R_{p,i} \Delta v_{i}^{2}$,\\
where $R_{p,i}$ and $\Delta v_{i}$ are the projected separation and relative line-of-sight velocity of the $i$th galaxy with respect to the group center, and $N$ is the number of group members.  
Although this estimator assumes virial equilibrium and isotropic velocities, it provides a reasonable first-order measure of the group mass. The inferred masses span $\sim10^{10}$–$10^{12}$\,\msun{} with a median $\log(M_d/M_\odot)=11.2$, typical for dwarf galaxy associations, highlighting their relatively low total gravitational potential compared to more massive galaxy groups and clusters.

\begin{figure}
\includegraphics[width=8.5cm]{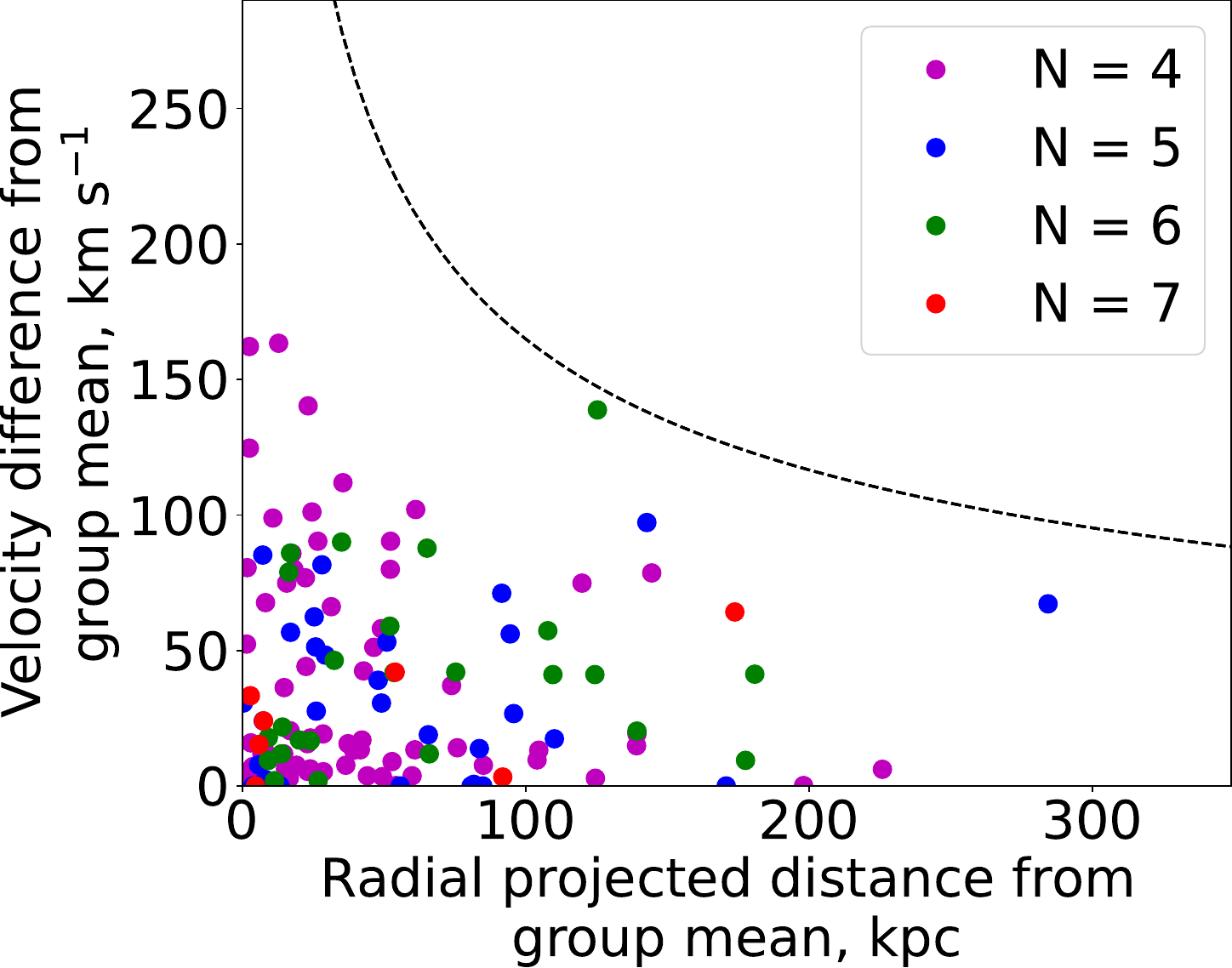}
\caption{Phase-space distribution of member galaxies of our groups. Different colors represent the group richness, as indicated in the legend: purple for richness 4, blue for 5, green for 6, and red for 7. The dashed curve denotes the escape velocity profile, calculated assuming a total group mass of 10$^{11.5}$\,\msun.}
\label{phase}
\end{figure}

In Figure~\ref{phase}, we show the projected phase-space distribution of our dwarf galaxy groups. The dashed curve marks the escape-velocity profile of a dark matter halo with a mass of $10^{11.5}\,M_\odot$. The fact that nearly all systems lie well within this boundary provides additional evidence that they are self-bound and reside within a common gravitational potential.
To further assess the dynamical state of the groups, we compute the three-dimensional velocity dispersion, $\sigma_{3D}$, by correcting the observed line-of-sight velocities for the projection effect. Assuming isotropy, we estimate $\sigma_{3D}$ using the relation:\\
$\sigma_{3D}$ = $\sqrt{3}$ $\times$ $\sqrt{\langle v^{2} \rangle - \langle v \rangle^{2}}$,\\
where $v$ is the line-of-sight velocity of each galaxy relative to the systemic velocity of its group. The resulting $\sigma_{3D}$ values span $\sim 50$--100~\kms, consistent with expectations for dynamically cold, low-mass systems.

Finally, we estimate the total mass-to-light ratios of the groups by dividing the dynamical mass by the total baryonic mass, taken as the sum of the stellar and \ion{H}{1} masses of all group members. These values should be regarded as upper limits. The available \ion{H}{1} data are incomplete---some galaxies lack detected \ion{H}{1} fluxes, and single-dish observations may miss low-level emission or blend contributions within the beam. Consequently, the total baryonic mass is likely underestimated, particularly in its neutral gas component, leading to an overestimation of the inferred mass-to-light ratios.

\begin{figure}
\includegraphics[width=8.5cm]{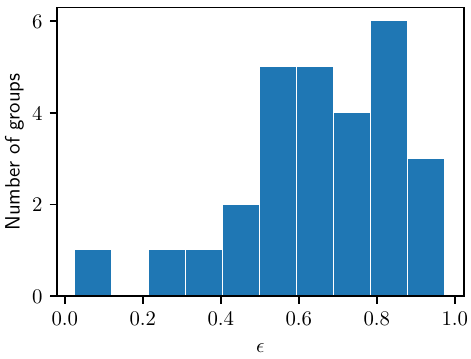}
\includegraphics[width=8.5cm]{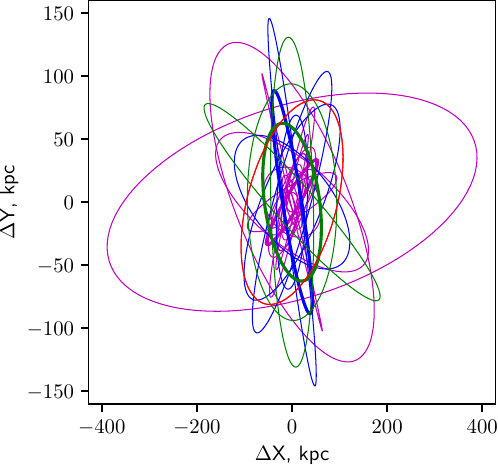}
\caption{Top: Distribution of ellipticity ($\epsilon$) of our dwarf galaxy groups, derived using principal component analysis. Bottom: Best-fit ellipses representing the spatial configuration of each group. The groups are categorized by their richness (i.e., number of member galaxies), with colors corresponding to those used in Figure \ref{phase}.}
\label{elip}
\end{figure}

In our previous work \citep{Paudel24}, we identified a dwarf galaxy group whose members are strikingly aligned along a nearly linear configuration, forming an unusually flattened, planar structure. To quantify such spatial arrangements across our full sample, we applied a Principal Component Analysis (PCA) to the projected positions of the group members, deriving a best-fit plane and computing the corresponding ellipticity ($\epsilon$) of each system. As shown in Figure~\ref{elip}, the resulting ellipticity distribution has a median value of $\epsilon = 0.67$, indicating that many of the groups exhibit markedly anisotropic, flattened configurations. In this context, larger $\epsilon$ values correspond to more elongated, planar structures.

\section{Discussion}

\begin{table}\centering
\caption{Number of identified groups for different group selection criteria.}
\label{gstab}
\begin{tabular}{|cc|c|c|c|c|}
\cline{1-6}
& & \multicolumn{4}{c|}{$R$ (kpc)}\\
\multirow{5}{*}{\rotatebox{90}{$\Delta V$ (km s$^{-1}$)}} & 
&150 & 200 & 250 & 300 \\ 
\cline{3-6}
&100& [\tre{14},\tgr{5},\tbl{3},0] & [\tre{12},\tgr{6},\tbl{3},1] & [\tre{13},\tgr{6},\tbl{3},1] & [\tre{12},\tgr{7},\tbl{3},1]\\[1ex]
\cline{2-6}
&150& [\tre{16},\tgr{6},\tbl{3},0] & [\tre{15},\tgr{5},\tbl{4},1] & [\tre{16},\tgr{5},\tbl{4},1] & [\tre{15},\tgr{6},\tbl{4},1]\\[1ex]
\cline{2-6}
&200& [\tre{18},\tgr{6},\tbl{3},0] & [\tre{17},\tgr{5},\tbl{4},1] & [\tre{18},\tgr{5},\tbl{4},1] & [\tre{17},\tgr{6},\tbl{4},1]\\[1ex]
\cline{2-6}
&250& [\tre{14},\tgr{5},\tbl{1},0] & [\tre{19},\tgr{6},\tbl{3},0] & [\tre{18},\tgr{5},\tbl{4},1] & [\tre{18},\tgr{6},\tbl{4},1]\\[1ex]
\cline{1-6}
\end{tabular}
\tablecomments{Each entry lists the number of groups with different richness levels. Colors indicate group richness: red = 4 members, green = 5 members, blue = 6 members, and black = 7 members.}
\end{table}

\subsection{Dynamical Consistency and Robustness of the Identified Groups}
In this work, we presented 28 dwarf-galaxy groups, each containing at least four confirmed members and residing in the local ($z < 0.02$) Universe. Although their full three-dimensional configurations cannot be directly measured, all systems are consistent with being gravitationally bound. Using our estimates of $\sigma_{3D}$, we computed the minimum total mass required for each group to remain bound by equating the escape velocity to $\sigma_{3D}$. We find that this minimum binding mass is typically about half of the total dynamical mass inferred from the projected mass estimator, providing an independent consistency check on the robustness of our mass estimates. The corresponding minimum total mass-to-light ratios span a range of $\sim 10$ to $110$.

To further assess whether the identified members are dynamically consistent with being gravitationally bound, we performed a series of Monte Carlo realizations in which galaxy velocities were perturbed according to the measured velocity dispersion while keeping their sky positions fixed. Assuming a characteristic halo mass of $10^{11.5}\,M_{\odot}$, we computed the corresponding escape velocity profile and evaluated the fraction of galaxies satisfying $|\Delta V| < V_{\rm esc}(R)$. We find that approximately $96\%$ of galaxies remain below the escape velocity boundary across the realizations. This result indicates that the vast majority of the identified systems are dynamically consistent with being gravitationally bound, suggesting that contamination from unbound interlopers is likely to be small. A comparable bound fraction ($\sim93\%$) is obtained when using the baseline selection criteria adopted in this work. We note, however, that this analysis does not explicitly account for projection effects in the observed phase space; therefore, the true fraction of unbound members may be somewhat higher.

To examine the sensitivity of our results to the adopted group selection criteria, we repeated the group identification using a range of projected separation and velocity thresholds. Specifically, we varied the projected radius from 150 to 300 kpc and the line-of-sight velocity difference from 100 to 250 km s$^{-1}$. For each combination of these parameters, we re-identified dwarf galaxy groups in the sample.  Across the explored parameter space, we identify up to 29 groups with at least four member galaxies\footnote{We selected 28 groups using our main selection criteria of $\Delta\,R < 300\,kpc$ and $\Delta_{v} < 200\,km\,s^{-1}$  }. The resulting group counts for each set of selection criteria are summarized in Table~\ref{gstab}. Overall, the number of detected groups varies only modestly across the tested thresholds, indicating that our results are not strongly sensitive to the exact choice of projected separation and velocity cuts.

\subsection{Compare to previous studies}

\begin{figure}
\includegraphics[width=8.5cm]{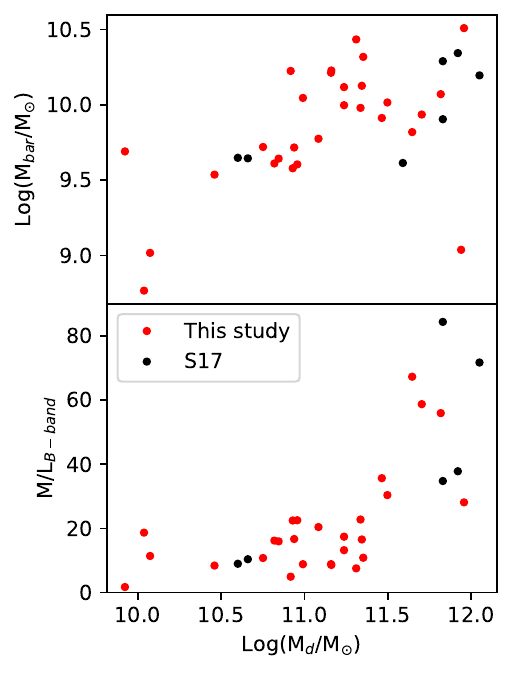}
\caption{Relation between baryonic mass and mass-to-light ratio with respect to the total dynamical mass. Red symbols represent our dwarf galaxy groups, while black symbols correspond to the compact groups studied in S17.}
\label{mll}
\end{figure}

\citet{Stierwalt17} (hereafter S17) investigated seven isolated compact groups of dwarf galaxies at $0.02 < z < 0.05$, corresponding to distances of $\sim 80$--200~Mpc. In contrast, our sample focuses on significantly closer systems, with a median distance of $\sim 55$~Mpc, placing them firmly within the local Universe. Among the S17 groups, only one system contains five members---the richest and most massive in their sample---with a total dynamical mass of $1.12\times10^{12}\,M_\odot$. The remaining systems consist of two quadruples and three triplets. By comparison, our sample includes one group with as many as seven members and three additional groups with six members, demonstrating that relatively rich, dwarf-only associations also exist in the nearby Universe.

\begin{figure}
\includegraphics[width=8.5cm]{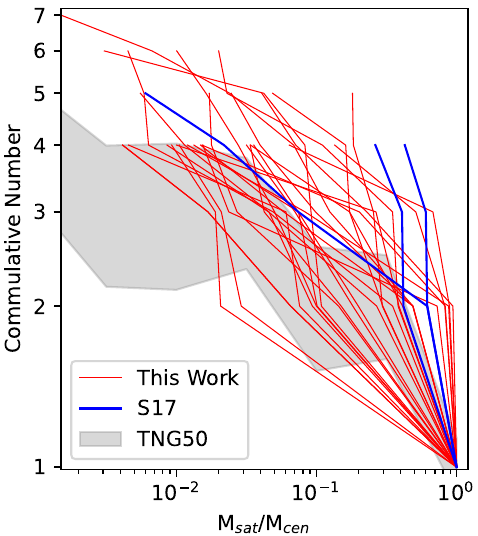}
\caption{Cumulative number of satellite galaxies as a function of the stellar mass ratio between centrals and their satellites. Red lines show our groups, the blue line represents the S17 groups (with four or more members), and the gray shaded region denotes the TNG50 prediction with $1\sigma$ scatter. Most observed groups fall within the TNG50 range, indicating strong agreement between simulations and observations.}
\label{tng}
\end{figure}

However, there are notable structural and dynamical differences between the two samples. The S17 groups are markedly more compact, with their largest system spanning only 80.5~kpc, whereas the projected sizes of our groups extend up to $\sim 180$~kpc. This contrast likely reflects the differing selection strategies: S17 intentionally targeted compact configurations, while our search adopted a broader spatial radius designed to capture gravitationally bound dwarf associations without imposing compactness as a prior.

Dynamically, the S17 groups also exhibit substantially higher internal velocity dispersions. Their most massive system (which has five members) shows a velocity dispersion exceeding 130~\kms. In contrast, our richest group, containing seven members, has a much lower dispersion of only 33~\kms, consistent with a dynamically colder configuration. These differences suggest that, although both samples consist of dwarf-dominated groups, they may represent distinct evolutionary stages or formation pathways.

In Figure~\ref{mll}, we compare the baryonic mass versus total dynamical mass (top panel) and the mass-to-light ratio versus total dynamical mass (bottom panel) for both samples. Despite the structural and dynamical differences noted above, our groups lie along the same baryonic--dynamical mass relation defined by the S17 systems. This consistency suggests that the overall dark matter fractions in the two samples are broadly comparable. It reinforces the idea that our more spatially extended, dynamically colder groups are likewise bound within dark matter halos, though at lower characteristic densities and velocity dispersions.

\subsection{Comparison to Simulations}

To compare our observational results with theoretical expectations, we use the halo catalog from the IllustrisTNG-50 cosmological simulation \citep{Pillepich19, Nelson19} at $z=0$. To ensure a fair comparison, we construct a mock-observed catalog by imposing the same stellar-mass completeness threshold derived for the data, restricting both simulated centrals and satellites to $\log(M_*/M_\odot) \geq 8.63$. This mass cut reproduces the SDSS selection function and removes low-mass subhalos that would not be detectable in the observational sample.


For each observed central galaxy, we randomly draw simulated centrals matched in stellar mass and select a corresponding number of satellites above the adopted completeness limit. We then derive the relation between the central-to-satellite stellar mass ratio and the cumulative satellite number (Figure~\ref{tng}). This Monte Carlo matching procedure is repeated 100 times to estimate statistical uncertainties and reduce sampling variance.

\trev{We find good agreement between the observations and TNG50 predictions at large mass ratios (i.e., for more massive satellites relative to their centrals). However, at lower mass ratios, a subset of high-richness systems exhibits a larger number of satellites than indicated by the TNG50 predictions. This discrepancy is partly driven by the relatively strict stellar mass threshold applied to the mock catalog, which is higher than the masses of some of the least massive satellites in the observed sample. As a result, low-mass satellites are likely undersampled in the mock catalogs, contributing to the observed difference. We further note that observational incompleteness in this context primarily refers to contamination from projection effects and interlopers, which can artificially increase the inferred group richness, rather than to missing low-mass galaxies. }

\section{Summary}

\begin{enumerate}
    \item We identify 28 dwarf galaxy groups in the local Universe ($z < 0.02$), each containing at least four confirmed members located within a projected radius of 300 kpc of the central galaxy and with line-of-sight velocity differences below 200\,km\,s$^{-1}$. The median redshift of the sample is $z = 0.0131$ (corresponding to a comoving distance of $\sim 55$\,Mpc), and the median stellar mass of the central galaxies is $5.3 \times 10^{9}$\,\msun, firmly placing them in the dwarf-galaxy regime.
    \item The majority of the groups exhibit low line-of-sight velocity dispersions, $\sigma_v < 50$\,\kms, consistent with dynamically cold and gravitationally bound systems. The inferred three-dimensional velocity dispersions span $\sigma_{3D} \sim 50$--100\,\kms.
    \item Our observational results show good agreement with the TNG50 predictions in terms of the cumulative number of satellite galaxies as a function of the stellar mass ratio between centrals and their satellites. This indicates that dwarf galaxies can host their own satellite systems embedded within extended dark matter halos, corroborating the hierarchical nature of structure formation in the $\Lambda$CDM framework.
\end{enumerate}

\newpage
\begin{acknowledgments}
SP and SJY acknowledge support from the Mid-career Researcher Program (RS-2023-00208957 and RS-2024-00344283, respectively) through Korea's National Research Foundation (NRF). 
SJY and CGS acknowledge support from the Basic Science Research Program (2022R1A6A1A03053472 and 2018R1A6A1A06024977, respectively) through Korea's NRF funded by the Ministry of Education. J.Y. was supported by a KIAS Individual Grant (QP089902) via the Quantum Universe Center at Korea Institute for Advanced Study

The DESI Legacy Imaging Surveys consist of three individual and complementary projects: the Dark Energy Camera Legacy Survey (DECaLS), the Beijing-Arizona Sky Survey (BASS), and the Mayall z-band Legacy Survey (MzLS). DECaLS, BASS and MzLS together include data obtained, respectively, at the Blanco telescope, Cerro Tololo Inter-American Observatory, NSF’s NOIRLab; the Bok telescope, Steward Observatory, University of Arizona; and the Mayall telescope, Kitt Peak National Observatory, NOIRLab. NOIRLab is operated by the Association of Universities for Research in Astronomy (AURA) under a cooperative agreement with the National Science Foundation. Pipeline processing and analyses of the data were supported by NOIRLab and the Lawrence Berkeley National Laboratory (LBNL). Legacy Surveys also uses data products from the Near-Earth Object Wide-field Infrared Survey Explorer (NEOWISE), a project of the Jet Propulsion Laboratory/California Institute of Technology, funded by the National Aeronautics and Space Administration. Legacy Surveys was supported by: the Director, Office of Science, Office of High Energy Physics of the U.S. Department of Energy; the National Energy Research Scientific Computing Center, a DOE Office of Science User Facility; the U.S. National Science Foundation, Division of Astronomical Sciences; the National Astronomical Observatories of China, the Chinese Academy of Sciences and the Chinese National Natural Science Foundation. LBNL is managed by the Regents of the University of California under contract to the U.S. Department of Energy. The complete acknowledgments can be found at https://www.legacysurvey.org/acknowledgment/.

This research used data obtained with the Dark Energy Spectroscopic Instrument (DESI). DESI construction and operations is managed by the Lawrence Berkeley National Laboratory. This material is based upon work supported by the U.S. Department of Energy, Office of Science, Office of High-Energy Physics, under Contract No. DE–AC02–05CH11231, and by the National Energy Research Scientific Computing Center, a DOE Office of Science User Facility under the same contract. Additional support for DESI was provided by the U.S. National Science Foundation (NSF), Division of Astronomical Sciences under Contract No. AST-0950945 to the NSF’s National Optical-Infrared Astronomy Research Laboratory; the Science and Technology Facilities Council of the United Kingdom; the Gordon and Betty Moore Foundation; the Heising-Simons Foundation; the French Alternative Energies and Atomic Energy Commission (CEA); the National Council of Science and Technology of Mexico (CONACYT); the Ministry of Science and Innovation of Spain (MICINN), and by the DESI Member Institutions: www.desi.lbl.gov/collaborating-institutions. The DESI collaboration is honored to be permitted to conduct scientific research on Iolkam Du’ag (Kitt Peak), a mountain with particular significance to the Tohono O’odham Nation. Any opinions, findings, and conclusions or recommendations expressed in this material are those of the author(s) and do not necessarily reflect the views of the U.S. National Science Foundation, the U.S. Department of Energy, or any of the listed funding agencies.

Funding for SDSS-III has been provided by the Alfred P. Sloan Foundation, the Participating Institutions, the National Science Foundation, and the U.S. Department of Energy Office of Science. The SDSS-III web site is http://www.sdss3.org/.

SDSS-III is managed by the Astrophysical Research Consortium for the Participating Institutions of the SDSS-III Collaboration including the University of Arizona, the Brazilian Participation Group, Brookhaven National Laboratory, Carnegie Mellon University, University of Florida, the French Participation Group, the German Participation Group, Harvard University, the Instituto de Astrofisica de Canarias, the Michigan State/Notre Dame/JINA Participation Group, Johns Hopkins University, Lawrence Berkeley National Laboratory, Max Planck Institute for Astrophysics, Max Planck Institute for Extraterrestrial Physics, New Mexico State University, New York University, Ohio State University, Pennsylvania State University, University of Portsmouth, Princeton University, the Spanish Participation Group, University of Tokyo, University of Utah, Vanderbilt University, University of Virginia, University of Washington, and Yale University.

\end{acknowledgments}

\facilities{DESI, SDSS, CDS, NED}
\software{The following software tools were utilized in this work: Astropy \cite{astropy}, Matplotlib \cite{matplotlib}, Numpy \cite{numpy}}


\section{Appendix}

\begin{figure*}[b]
\includegraphics[page=1]{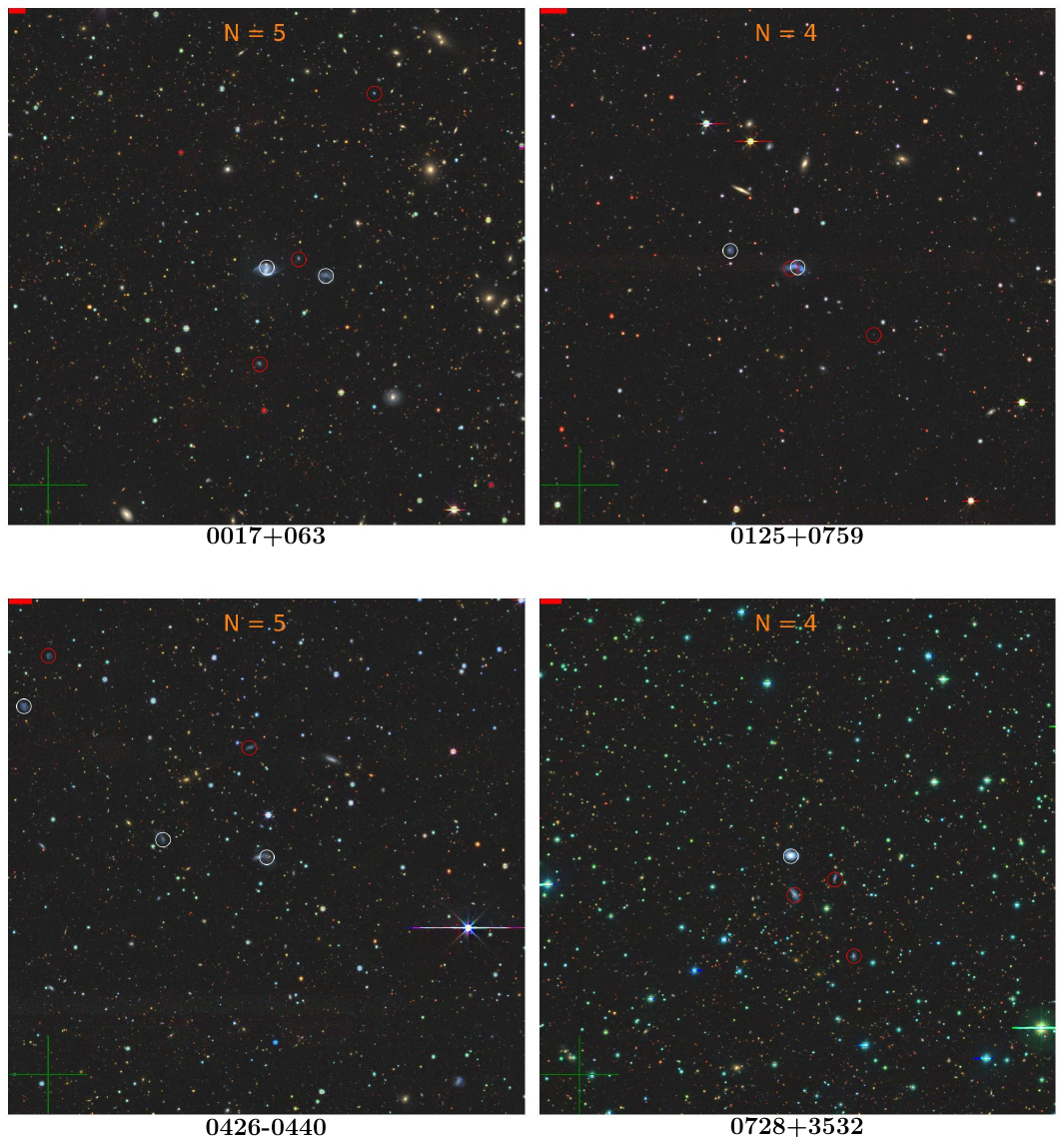}
\caption{Similar to Figure \ref{lgrp}. The group IDs are listed at the bottom of the figure. }
 \label{apfig}
\end{figure*}

Figure \ref{apfig} shows postage images as shown in Figure \ref{lgrp} for all other groups.

\includegraphics[page=2]{mkfig1}
\newpage
\includegraphics[page=3]{mkfig1}
\newpage
\includegraphics[page=4]{mkfig1}
\newpage
\includegraphics[page=5]{mkfig1}
\newpage
\includegraphics[page=6]{mkfig1}
\newpage
\includegraphics[page=7]{mkfig1}

\end{document}